\documentstyle[psfig]{mn}
\newif\ifAMStwofonts

\def\simlt{\lower.5ex\hbox{$\; \buildrel < \over \sim \;$}}
\def\simgt{\lower.5ex\hbox{$\; \buildrel > \over \sim \;$}}

\ifoldfss
  \ifCUPmtlplainloaded \else
    \NewTextAlphabet{textbfit} {cmbxti10} {}
    \NewTextAlphabet{textbfss} {cmssbx10} {}
    \NewMathAlphabet{mathbfit} {cmbxti10} {} 
    \NewMathAlphabet{mathbfss} {cmssbx10} {} 
  \fi
  \ifAMStwofonts
    \ifCUPmtlplainloaded \else
      \NewSymbolFont{upmath} {eurm10}
      \NewSymbolFont{AMSa} {msam10}
      \NewMathSymbol{\upi}     {0}{upmath}{19}
      \NewMathSymbol{\umu}     {0}{upmath}{16}
      \NewMathSymbol{\upartial}{0}{upmath}{40}
      \NewMathSymbol{\leqslant}{3}{AMSa}{36}
      \NewMathSymbol{\geqslant}{3}{AMSa}{3E}

      \let\leq=\leqslant 
      \let\geq=\geqslant 
    \fi
  \fi
\fi 

\ifnfssone
  \newmathalphabet{\mathit}
  \addtoversion{normal}{\mathit}{cmr}{m}{it}
  \addtoversion{bold}{\mathit}{cmr}{bx}{it}
  \newmathalphabet{\mathbfit} 
  \addtoversion{normal}{\mathbfit}{cmr}{bx}{it}
  \addtoversion{bold}{\mathbfit}{cmr}{bx}{it}
  \newmathalphabet{\mathbfss} 
  \addtoversion{normal}{\mathbfss}{cmss}{bx}{n}
  \addtoversion{bold}{\mathbfss}{cmss}{bx}{n}
  \ifAMStwofonts
    \ifCUPmtlplainloaded \else
      \UseAMStwoboldmath
      \makeatletter
      \new@mathgroup\upmath@group
      \define@mathgroup\mv@normal\upmath@group{eur}{m}{n}
      \define@mathgroup\mv@bold\upmath@group{eur}{b}{n}
      \edef\UPM{\hexnumber\upmath@group}
      \new@mathgroup\amsa@group
      \define@mathgroup\mv@normal\amsa@group{msa}{m}{n}
      \define@mathgroup\mv@bold\amsa@group{msa}{m}{n}
      \edef\AMSa{\hexnumber\amsa@group}
      \makeatother
      \mathchardef\upi="0\UPM19
      \mathchardef\umu="0\UPM16
      \mathchardef\upartial="0\UPM40
      \mathchardef\leqslant="3\AMSa36
      \mathchardef\geqslant="3\AMSa3E

      \let\leq=\leqslant 
      \let\geq=\geqslant 
    \fi
  \fi
\fi 

\ifnfsstwo
  \DeclareMathAlphabet{\mathbfit}{OT1}{cmr}{bx}{it}
  \SetMathAlphabet\mathbfit{bold}{OT1}{cmr}{bx}{it}
  \DeclareMathAlphabet{\mathbfss}{OT1}{cmss}{bx}{n}
  \SetMathAlphabet\mathbfss{bold}{OT1}{cmss}{bx}{n}
  \ifAMStwofonts
    \ifCUPmtlplainloaded \else
      \DeclareSymbolFont{UPM}{U}{eur}{m}{n}
      \SetSymbolFont{UPM}{bold}{U}{eur}{b}{n}
      \DeclareSymbolFont{AMSa}{U}{msa}{m}{n}
      \DeclareMathSymbol{\upi}{0}{UPM}{"19}
      \DeclareMathSymbol{\umu}{0}{UPM}{"16}
      \DeclareMathSymbol{\upartial}{0}{UPM}{"40}
      \DeclareMathSymbol{\leqslant}{3}{AMSa}{"36}
      \DeclareMathSymbol{\geqslant}{3}{AMSa}{"3E}

      \let\leq=\leqslant 
      \let\geq=\geqslant 
    \fi
  \fi
\fi 

\ifCUPmtlplainloaded \else
  \ifAMStwofonts \else 
    \def\upi{\pi}
    \def\umu{\mu}
    \def\upartial{\partial}
  \fi
\fi

\title [Spatially-resolved spectra of UU Aquarii]
		{Spatially-resolved spectra of the accretion disc of the novalike
		UU Aquarii}
\author[R. Baptista et~al.]
	{Raymundo Baptista$^1$, C. Silveira$^1$, J.E. Steiner$^2$
	and Keith Horne$^3$ \\
	$^1$ Departamento de F\'\i sica, Universidade Federal de Santa Catarina,
    Campus Trindade, 88040-900, Florian\'opolis - SC, Brazil, \\
    ~ email: bap@fsc.ufsc.br, silveira@fsc.ufsc.br \\
	$^2$ Laborat\'orio Nacional de Astrof\'\i sica-LNA/CNPq, CP 21, 37500-000,
	Itajub\'a, Brazil, email: steiner@lna.br \\
	$^3$ School of Physics \& Astronomy, University of St.\,Andrews,
	North Haugh, St.\,Andrews, Fife, KY16 9SS, Scotland, \\
    ~ email: kdh1@st-and.ac.uk \\
	}
\date{Accepted for publication at Monthly Notices of the Royal Astronomical
		Society}

\pagerange{1-13}
\pubyear{2000}

\begin{document}

\maketitle

\begin{abstract}

Time-resolved spectroscopy of the novalike variable UU Aquarii is analyzed
with eclipse mapping techniques to produce spatially resolved spectra
of its accretion disc and gas stream as a function of distance from disc
centre in the range 3600-6900 \AA.  The spatially-resolved spectra show
that the continuum emission becomes progressively fainter and redder for
increasing disc radius -- reflecting the radial temperature gradient --
and reveal that the H\,I and He\,I lines appear as deep, narrow absorption
features in the inner disc regions transitioning to emission with P Cyg
profiles for intermediate and large disc radii.  The spectrum of the
uneclipsed component has strong H\,I and He\,I emission lines plus a Balmer
jump in emission and is explained as optically thin emission from a
vertically extended disc chromosphere + wind. 
Most of the line emission probably arises from the wind.
The spatially-resolved spectra also suggest the existence of gas stream 
``disk-skimming'' overflow in UU~Aqr, which can be seen down to
$R\simeq 0.2\; R_{L1}$. The comparison of our eclipse maps with those 
of Baptista, Steiner \& Horne (1996) suggests that the asymmetric structure 
in the outer disc previously identified as the bright spot may be the 
signature of an elliptical disc similar to those possibly present in 
SU~UMa stars during superoutbursts.

\end{abstract}

\begin{keywords}
binaries: close -- novae, cataclysmic variables -- eclipses -- accretion
discs -- stars: individual: (UU Aquarii).
\end{keywords}

\section{Introduction}

The standard picture of a novalike system is that of a close binary in
which a late type star fills its Roche lobe and transfers matter to a
companion white dwarf via an accretion disc. A bright spot is expected 
to form where the gas stream from the donor star hits the edge of the
accretion disc.

The SW Sex stars (Thorstensen et~al. 1991) form a sub-class of the
novalikes with orbital periods in the range 3-4 hs that do not seem to
fit within the above standard picture, displaying a range of peculiarities:
(1) single peaked asymmetric emission lines showing little eclipse, 
(2) large ($\sim 70\degr$) phase shifts between photometric and
spectroscopic conjunction, 
(3) orbital phase-dependent absorption in the Balmer lines,
(4) Doppler tomograms bright in the lower-left quadrant with small
or no sign of disc emission, and
(5) v-shaped continuum eclipses implying in flat radial temperature 
profiles in the inner disc (e.g., Warner 1995 and references therein).
Earlier proposals to explain the phenomenon include accretion disc winds
(Honeycutt, Schlegel \& Kaitchuck 1986), magnetic white dwarfs
disrupting the inner disc (Williams 1989), and gas stream overflow
(Hellier \& Robinson 1994). The two most recent models proposed to
explain the phenomenology of the SW Sex stars are the disc-anchored
magnetic propeller (Horne 1999) and a combination of stream overflow
+ disc winds (Hellier 1999).

UU~Aqr is an eclipsing novalike (P$_{\rm orb}= 3.9$\,hr) whose 
spectrum is dominated by single-peaked strong Balmer and He\,I emission
lines (e.g., Downes \& Keyes 1988). H$\alpha$ spectroscopy revealed
that the line profile is highly asymmetric and phase-dependent 
and that the spectroscopic conjunction lags mid-eclipse
by $\sim 0.15$ cycle (Haefner 1989; Diaz \& Steiner 1991).
The lack of the rotational disturbance typical of emitting accretion
discs during eclipse in H$\beta$ led Hessman (1990) to the suggestion
that the emission lines have a non-disc origin.

Baptista, Steiner \& Cieslinski (1994; hereafter BSC94) derived a
photometric model for the binary with $q= 0.30$, M$_1 = 0.67\:
{\rm M}_{\odot}$, an inclination of $i= 78$ degrees.
From the analysis of mid-eclipse fluxes they suggested that the Balmer
lines are formed in an extended region only partially occulted during
eclipse, possibly in a wind emanating from the inner disc.
They also found that UU Aqr presents long-term brightness variations of
low amplitude ($\simeq 0.3$ mags) on timescales of years.

The eclipse mapping study of Baptista, Steiner \& Horne (1996;
thereafter BSH96) indicates that the inner disc of UU Aqr is
optically thick, resulting in a distance estimate of 200 pc.
Temperatures in the disc range from $\sim 6000$\,K in the outer 
regions to $\sim 16000$\,K near the white dwarf at disc centre.
The radial temperature profiles in the high state follow the
T$\:\propto R^{-3/4}$ law in the outer and intermediate disc regions
but flattens off in the inner disc, leading to mass accretion rates
of $10^{-9.2}\:{\rm M_{\odot}\,yr}^{-1}$ at $R= 0.1\: R_{\rm L1}$
and $10^{-8.8}\:{\rm M_{\odot}\,yr}^{-1}$ at $R= 0.3\: R_{\rm L1}$
($R_{\rm L1}$ is the distance from disc centre to the inner Lagrangian
point). Together with other characteristics, this led BSH96 to suggest
that UU Aqr was an SW Sex star.
The comparison of eclipse maps of the low and high states revealed that
the differences are due to changes in the structure of the outer parts
of the disc, the most noticeable effect being the appearance of a
conspicuous red, bright structure at disc rim, which the authors
identified with the bright spot.

According to BSH96, the \.{M} of UU Aqr is barely above the critical
limit for disc instability to set in. Warner (1997) noted that the
outer disc temperature is only 6000 K and remarked that small variations
in \.{M} could lead to dwarf novae type outbursts.
Honeycutt, Robertson \& Turner (1998) performed a long-term photometric
monitoring of UU Aqr which confirmed the high and low brightness states
of BSC94 and revealed the existence of small amplitude ($\simlt 1.0$ mag)
brightness variations on timescales of a few days, which they called
`stunted outbursts'.

The detailed spectroscopic study of Hoard et~al. (1998) reinforced the
classification of UU Aqr as an SW Sex star.
They found evidences for the presence of a bright spot at the impact site
of the gas stream with the edge of the disc, and a non-axisymmetric,
vertically and azimuthally extended absorbing structure in the disc.
They proposed an explanation for the absorbing structure as well as for
the other spectroscopic features of UU Aqr in terms of the explosive
impact of the accretion stream with the disc.
Optical and ultraviolet spectroscopy by Kaitchuck et~al. (1998)
shows a secondary eclipse at phase 0.4 in the optical and Balmer lines
(but not in the UV continuum or lines) which they suggested may be
caused by an occultation of the bright spot and stream region by material
suspended above the inner disc.

In this paper we report on the analysis of time-resolved spectroscopy
of UU Aqr with multi-wavelength eclipse mapping techniques to derive 
spatially-resolved spectra of the accretion flow in this binary.
Section~\ref{data} describes the observations and data reduction procedures,
while section~\ref{analise} describes the analysis of the light curves with
eclipse mapping techniques.
Section~\ref{resultados} presents eclipse maps at selected wavelengths,
the radial intensity and brightness temperature distributions,
spatially resolved spectra of the accretion disc and gas stream as well
as the spectrum of the uneclipsed component.
The results are discussed in section~\ref{discussao} and summarized in section~\ref{conclusao}.

\section{Observations} \label{data}

Time-resolved spectroscopy covering 5 eclipses of UU~Aqr was obtained
with the 2.1-m telescope at the Kitt Peak National Observatory (KPNO) on
July-August 1993 in the spectral range 3500--6900 \AA\ (spectral
resolution of $\Delta\lambda= 1.5$ \AA\ pixel$^{-1}$).
The observations consist of 5 sets of $\simeq 100$ short exposure
($\Delta t=30s$) spectra at a time resolution of 50\,s. A close comparison
star (star C1 of BSC94) was included in the slit to allow correction of sky
transparency variations and slit losses. The observations (summarized in
Table~\ref{tab1}) were performed under good (cloud-free) sky conditions and
at small to moderate air masses ($X \leq 1.4$) except for run 1, which
started while the object was still at a reasonably high zenith angle
($X= 2.2$).
%
\begin{table*}
 \centering
 \begin{minipage}{120mm}
  \caption{Journal of the observations.}
  \label{tab1}
\begin{tabular}{@{}lccccccc@{}}
~~Date & Run & \multicolumn{2}{c}{UT} & No. of & Spectral & Cycle &
Phase range \\ [-0.5ex]
~(1993) && start & end & spectra & range (\AA) & number & (cycle) \\ [1ex]
22 July   & 1 & 05:56 & 07:33 & 105 & 3564.0--6766.5 & 17389 & $-0.20,+0.21$ \\
27 July   & 2 & 07:59 & 09:40 & 111 & 3601.6--6805.5 & 17420 & $-0.11,+0.32$ \\
13 August & 3 & 07:52 & 09:25 & 101 & 3646.5--6850.5 & 17524 & $-0.22,+0.17$ \\ 
15 August & 4 & 07:01 & 08:37 & 106 & 3649.5--6853.5 & 17536 & $-0.20,+0.20$ \\
16 August & 5 & 06:18 & 08:18 & 110 & 3649.5--6853.5 & 17542 & $-0.27,+0.24$ \\ 
\end{tabular}
\end{minipage}
\end{table*}

The data were bias-subtracted and corrected for flat-field and slit
illumination effects using standard {\sc iraf} procedures.
1-D spectra of both variable and comparison star were extracted
with the optimal extraction algorithm of Horne (1986). 
The individual spectra were checked for the presence of possible cosmic
rays and, when appropriate, were corrected by interpolation from the
neighboring wavelengths. Arc-lamp observations were used to calibrate the
wavelength scale (accuracy of 0.15 \AA). Observations of the standard spectrophotometric stars
BD+28\,4211 and G191\,B2B (Massey et~al. 1988) were used to derive the
instrumental sensitivity function and to flux calibrate the set of extracted
spectra on each night. Error bars were computed taking into account the photon
count noise and the sensitivity response of the instrument.

The reduced spectra were combined to produce trailed spectrograms of the
variable and the comparison star for each night. The display of the trailed
spectrograms of the comparison star shows that there were non negligible sky
transparency variations and/or time-dependent slit losses along the runs.
We defined a reference spectrum of the comparison star by computing an average
of 40 spectra on night 5 corresponding to the time for which the star was
closest to zenith.
We normalized the spectrograms of the comparison star by dividing each
spectrum by the reference spectrum. A 2-D cubic spline fit was used to
produce a smoothed version of the normalized spectrograms.
The sky transparency variations and variable slit losses were corrected
by dividing the spectrogram of the variable by the smoothed, normalized
spectrogram of the comparison star on each night (a procedure analogous
to the flat-field correction).
The reference spectrum is consistent with the UBVRI photometry of star C1
(BSC94) at the 1-$\sigma$ level. Therefore, the absolute photometric accuracy
of these observations should be better than 10 per cent.

\begin{figure}
\centerline{\psfig{figure=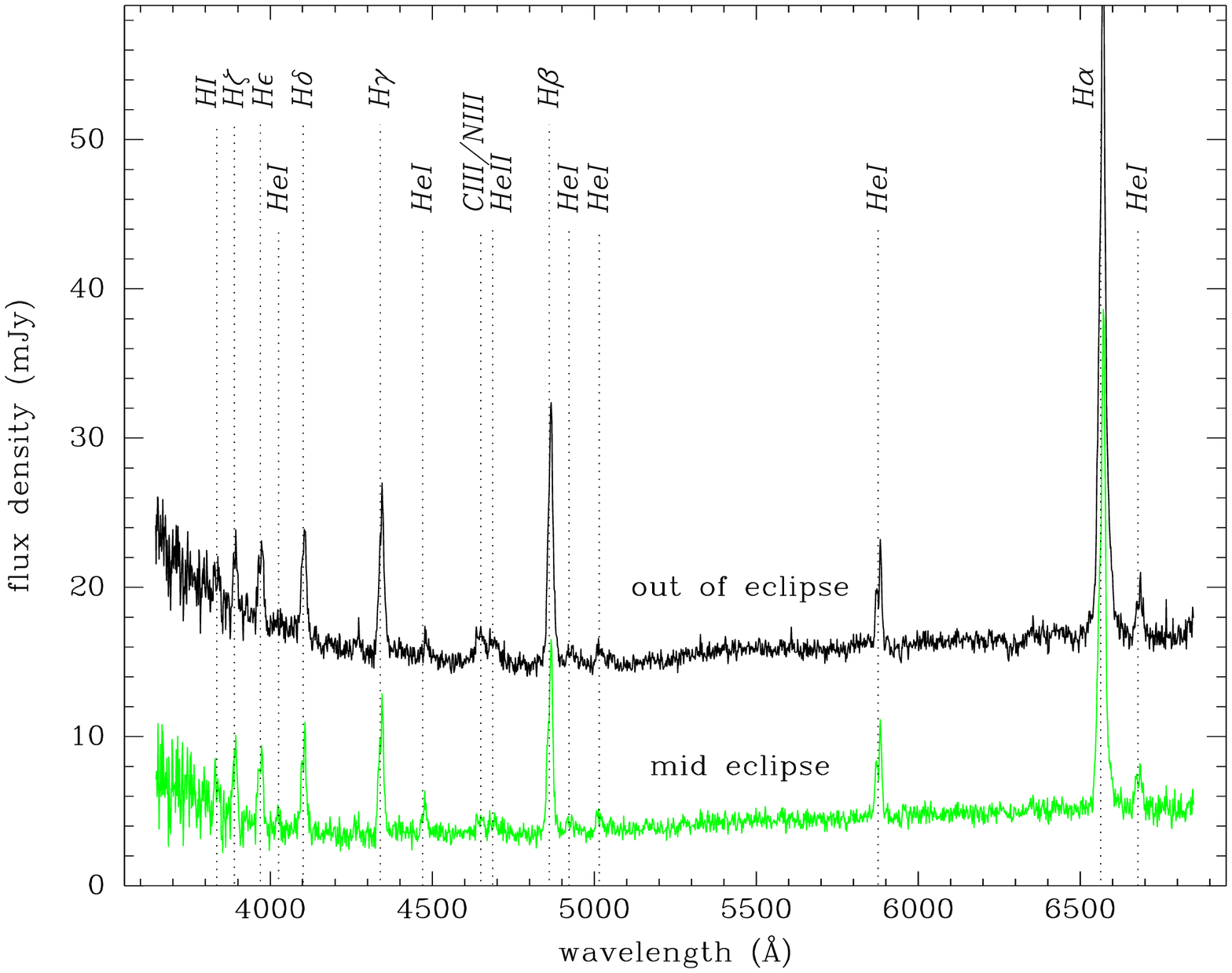,angle=-90,width=10.5cm,rheight=7.5cm}}
\caption{Average out-of-eclipse (black, phase range $-0.22$ to $-0.06$ cycle)
and mid-eclipse (light gray, phase range $-0.025$ to 0.025 cycle) spectra of
UU~Aqr on 1995 August 13. Major emission features are indicated by vertical
dotted lines.}
\label{fig1}
\end{figure}

Fig.\,\ref{fig1} shows average out-of-eclipse and mid-eclipse spectra of
UU~Aqr on 1995 August 13. The spectra are dominated by strong single-peaked
Balmer emission lines but also show He\,I lines and the blend of
C\,III, N\,III and He\,II lines at $\sim 4650$ \AA. The emission lines
have asymmetrical shapes, the red side of the line being stronger --
in accordance with the results of Hessman (1990) and Diaz \& Steiner (1991).
The He\,I lines and the higher energy Balmer lines show a possible double
peak structure suggesting either classical double-peaked emission from a
highly inclined disc or single peaked emission with a central absorption
component. While the continuum is reduced by a factor
$\simeq 3$ during eclipse, the emission lines
suffer a much smaller reduction in flux suggesting that they possibly
arise from a vertically-extended source larger than the accretion disc
(responsible for the continuum emission), in accordance with inferences
drawn by BSC94.
%
%

\begin{figure}
\centerline{\psfig{figure=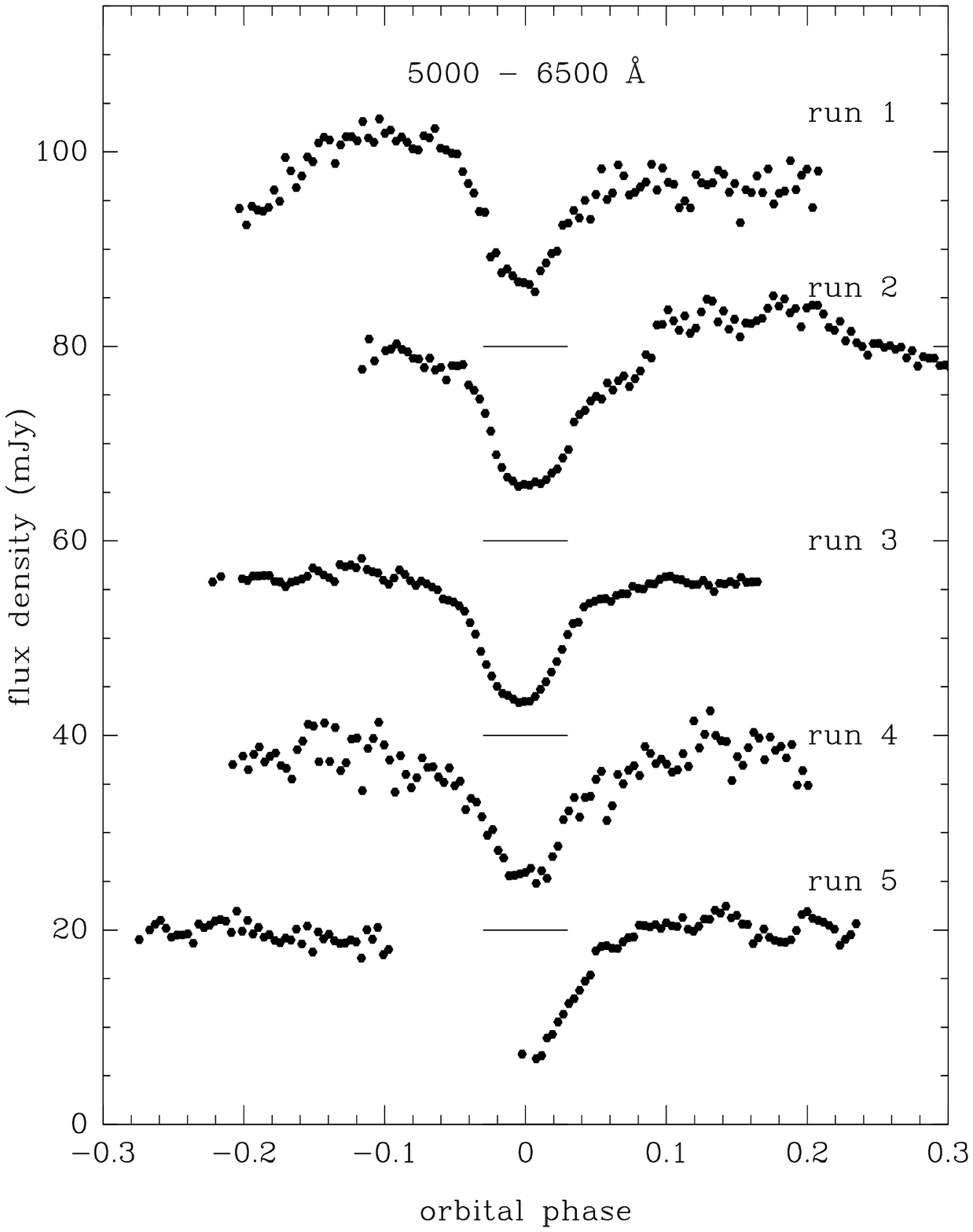,width=10cm,rheight=12cm}}
\caption{Broad band lightcurves of the UU Aqr dataset. The curves are
progressively displaced upwards by 20 mJy for visualization purposes.
Horizontal lines at mid-eclipse show the true zero level in each case.}
\label{fig2}
\end{figure}

Fig.\,\ref{fig2} shows lightcurves of the 5 runs in the broad band $5000-
6500$ \AA. The gap in run 5 is due to
an interruption of the observations to check the telescope focus.
The lightcurves have similar eclipse shapes and out of eclipse flux levels,
with variations at the level of $\simlt 20$\ per cent between the runs.
Indications that the observations were performed while UU~Aqr was in its
high brightness state come from the eclipse shape and average out of eclipse
flux level. The latter suggests that the object was even slightly brighter
than the typical high brightness state of BSC94.
These remarks are in agreement with the historical lightcurve of Honeycutt
et~al. (1998, see their fig.\,1), which shows that UU~Aqr reached a maximum
of its long-term average brightness level during 1993, the epoch of our
observations. The spectral range of the lightcurves in Fig.\,1 corresponds
roughly to the $V$ band.
The average out of eclipse level of all runs yields an approximate mean
magnitude of $V= 13.2 \pm 0.2$ mag, consistent with the value drawn from
the lightcurve of Honeycutt et~al. (1998), of $V= 13.4 \pm 0.6$ mag.
%
%

\section{Data analysis} \label{analise}

\subsection{Light curve construction}

The spectra were divided into 226 passbands of 15 \AA\ in the continuum and
fainter lines, and $\simeq 500\; km\,s^{-1}$ across the most prominent lines.
For each passband a lightcurve was extracted by computing the average flux
on the corresponding wavelength range and phase folding the resulting data
according to the ephemeris of BSC94. A phase correction of $-0.003$ cycle was
further applied to the data to make the centre of the white dwarf eclipse
coincident with phase zero.  For those passbands including emission lines
the light curves comprise the total flux at the corresponding bin with
no subtraction of a possible underlying continuum contribution.

Since the dataset correspond to the same brightness level it was possible
to combine the lightcurves of all runs to produce average lightcurves for
each passband. This is helpful to increase the signal-to-noise ratio of the
lightcurves and to reduce the influence of flickering in the eclipse shape.
For each passband, we first normalized the individual lightcurves by
fitting a spline function to the phases outside eclipse and dividing the
lightcurve by the fitted spline. The normalized lightcurves were combined by
separating the data into phase bins of 0.0038~cycle and computing the median
for each bin. The median of the absolute deviations with respect to the
median is taken as the corresponding uncertainty.
The resulting lightcurve is scaled back to flux units by multiplying the
combined lightcurve by the median flux of the spline functions at phase zero.
This procedure removes orbital variations outside eclipse with
only minor effects on the eclipse shape itself.

\subsection{Eclipse mapping}
\label{mem}

The eclipse mapping method was used to solve for a map of the disc
brightness distribution and for the flux of an additional uneclipsed
component in each passband.  For the details of the method the reader
is referred to Horne (1985, 1993), Baptista \& Steiner (1993) and
Rutten et al. (1994).

For our analysis we adopted the same eclipse map of BSH96, a $51 \times 51$
pixel grid centred on the primary star with side $2 \:R_{L1}$ where
$R_{L1}$ is the distance from the disc centre to the inner Lagrangian point.
This choice provides maps with a nominal spatial resolution of $0.039 \:
R_{L1}$, comparable to the expected size of the white dwarf in UU~Aqr
($\simeq 0.032\: R_{L1}$). The eclipse geometry is specified by the mass
ratio $q$ and the inclination $i$. We adopted the parameters of BSC94,
$i= 78\degr$ and $q=0.3$.
The specific intensities in the eclipse map were computed assuming
R$_{L1}= 0.74 \; R_\odot$ (BSC94) and a distance of 200 pc (BSH96).

The statistical uncertainties of the eclipse maps were estimated with a
Monte Carlo procedure (e.g., Rutten et al. 1992; Baptista et~al. 1995).
For a given narrow-band lightcurve a set of 10 artificial lightcurves is
generated, in which the data points are independently and randomly
varied according to a Gaussian distribution with standard deviation
equal to the uncertainty at that point.  The lightcurves are fitted
with the eclipse mapping algorithm to produce a set of randomized
eclipse maps. These are combined to produce an average map and a map
of the residuals with respect to the average, which yields the statistical
uncertainty at each pixel.
The uncertainties obtained with this procedure will be used when
estimating the errors in the derived radial temperature and intensity
profiles as well as in the spatially-resolved spectra.

Average light curves, fitted models, and eclipse maps at selected passbands
are show in Figs.\,\ref{fig3} and \ref{fig4}. These will be discussed in
detail in section\,\ref{resultados}.

\section{Results} \label{resultados}

\subsection{Accretion disc structure}
\label{estrutura}

In this section we compare eclipse maps at selected passbands in order
to study the structure of the accretion disc at different wavelengths.

\begin{figure}
\centerline{\psfig{figure=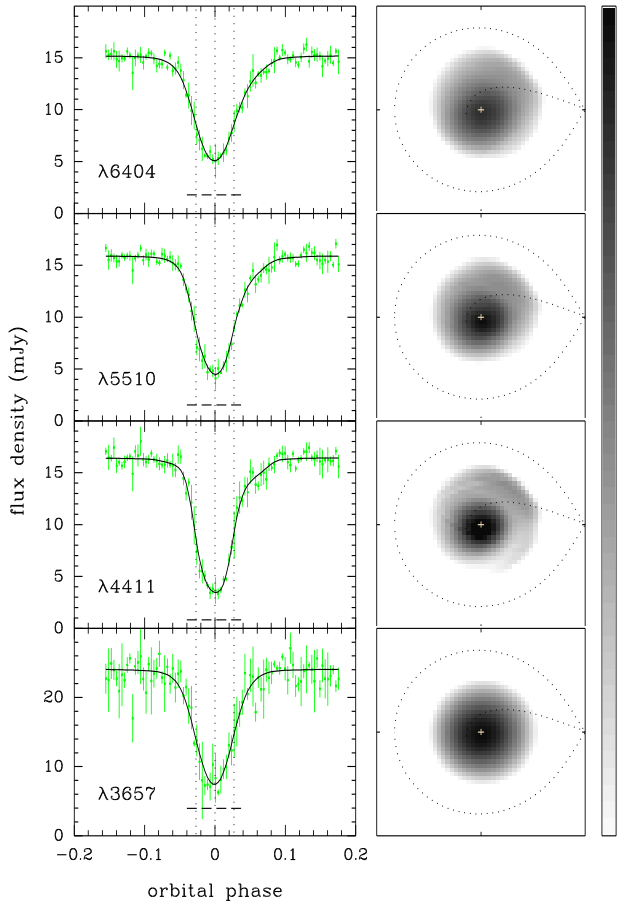,width=11cm,rheight=13.5cm}}
\caption{ Lightcurves (left) and eclipse maps (right) at selected continuum
	passbands. Data lightcurves are shown as gray dots with error bars and
	the fitted models appear as solid black lines. A horizontal dashed line
	depicts the uneclipsed component in each case. Labels indicate the
	central wavelength of each passband. Eclipse maps are shown to the right
	in a logarithmic grayscale: dark regions are brighter; white corresponds
	to $\log I_\nu= -6.5$, and black to $\log I_\nu= -3.1$. Dotted curves
	show the projection of the primary Roche lobe onto the orbital plane
	and the theoretical gas stream trajectory; the secondary star is to the
	right of each panel and the stars rotate counter-clockwise. }
\label{fig3}
\end{figure}

Fig.\,\ref{fig3} shows lightcurves (left panels) and eclipse maps (right
panels) of 4 selected continuum passbands close to the Johnson-Cousins
UBVR effective wavelengths in order to allow a comparison with
the results of BSH96.
Dashed horizontal lines depict the uneclipsed component in each case.
The continuum lightcurves show a deep eclipse with a slightly asymmetric
egress shoulder which is more pronounced for longer wavelengths.
This results in eclipse maps with brightness distributions concentrated
towards disc centre and asymmetric structures in the trailing quadrant of
the disc closest to the secondary star (the upper right quadrant in the
eclipse maps of Fig.\,\ref{fig3}).
The uneclipsed component at $\lambda 3657$ is perceptibly 
larger than at $\lambda 4411$, suggesting that the Balmer jump is
in emission and that the uneclipsed light has an important contribution
from optically thin gas. This is in line with previous results by
BSC94 and BSH96.
The eclipse shapes and out of eclipse levels resemble those
of the high brightness state observed by BSH96, although with a less
pronounced asymmetry at eclipse egress. Accordingly, the eclipse maps
clearly lack the noticeable asymmetric structure at disc edge which was
the main characteristic of the high state (BSH96, see their Fig.\,3).
We will return to this point in section\,\ref{discussao}.
%
%

\begin{figure}
\centerline{\psfig{figure=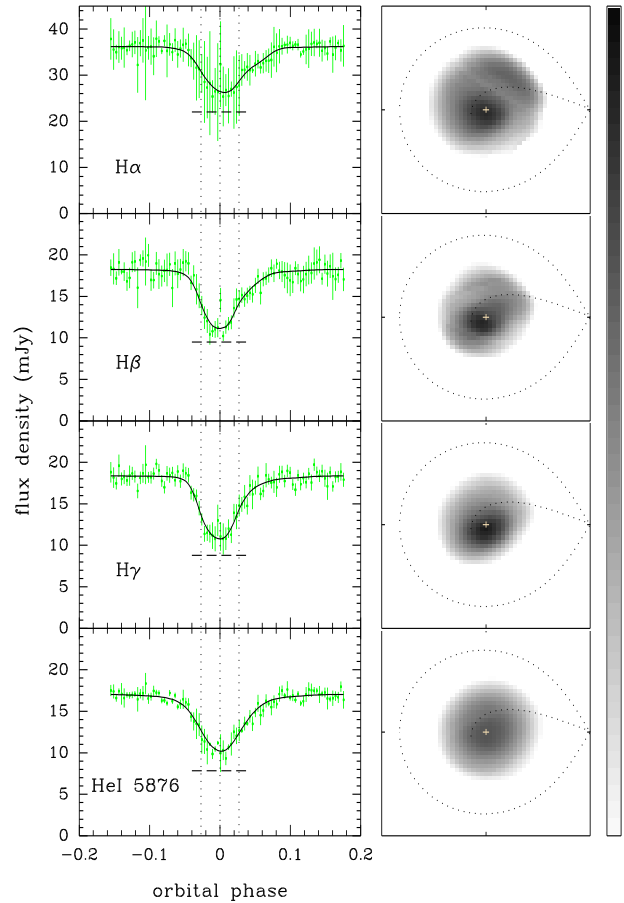,width=11cm,rheight=13.5cm}}
\caption{ Lightcurves (left) and eclipse maps (right) for the H$\alpha$,
	H$\beta$, H$\gamma$ and He\,I $\lambda 5876$ line centre passbands.
	The notation and logarithmic grayscale are the same as in
	figure\,\ref{fig3}. }
\label{fig4}
\end{figure}

Fig.\,\ref{fig4} shows lightcurves and eclipse maps for the line centre
passbands of H$\alpha$, H$\beta$, H$\gamma$ and He\,I $\lambda$5876.
We remark that the line lightcurves include the total flux at the 
corresponding wavelength range with no subtraction of an interpolated
continuum. The eclipses are shallow, leading to brightness distributions
which are flatter than those of the continuum. 
Similar to the continuum maps, the asymmetry in the egress shoulder is
more pronounced for the lines at longer wavelengths.
The uneclipsed components are considerably larger than in the continuum,
indicating that the uneclipsed spectrum has strong Balmer and He\,I
emission lines.
The large error bars of the H$\alpha$ centre lightcurve is due not to
low signal-to-noise ratio but to the variability of the eclipse shape at
this wavelength. This effect is also seen, although to a lesser extent,
in H$\beta$ and H$\gamma$.
%
%

\begin{figure}
\centerline{\psfig{figure=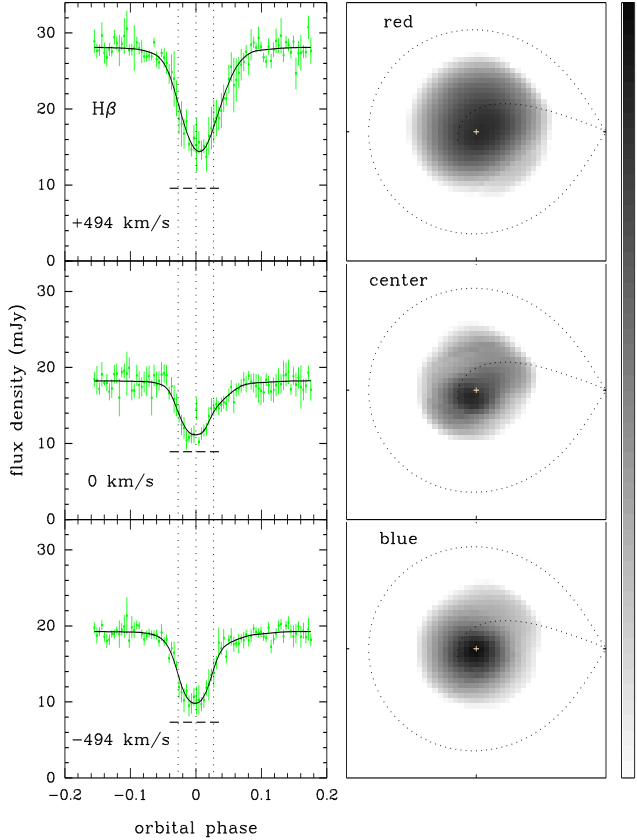,width=10cm,rheight=12.2cm}}
\caption{ H$\beta$ velocity-resolved lightcurves (left) and eclipse maps
	(right) at velocities of $-494,\; 0,\; {\rm and}\; +494	\;km \;s^{-1}$.
	The notation and logarithmic grayscale are the same as in
	figure\,\ref{fig3}. }
\label{fig5}
\end{figure}

Fig.\,\ref{fig5} shows (Doppler) velocity-resolved lightcurves (left) and
eclipse maps (right) across the H$\beta$ line.
There is marginal evidence of rotational disturbance: the minimum of the
blue bin lightcurve ($- 494 \; km \;s^{-1}$) is slightly displaced towards 
negative phases while that of the red bin lightcurve ($+ 494 \; km \;s^{-1}$)
is correspondingly displaced towards positive phases, suggesting that the
line emitting gas rotates in the prograde sense.
However, the eclipse maps in the symmetric velocity bins do not show the
mirror symmetry (over the line joining both stars) expected for line
emission from a Keplerian disc around the white dwarf.
Equally remarkable are the facts that the lightcurve in the red bin has a
much larger out-of-eclipse flux than its blue counterpart and that the
corresponding eclipse map is perceptibly brighter than that of the blue
bin anywhere. A similar behaviour is found in the other lines for which
velocity-resolved maps were obtained.
This cannot be attributed to the underlying continuum since the
interpolated continuum has essentially a constant level across each line.
It seems clear that most of the line emission does not arise from a
disc in Keplerian rotation.
%
%

\subsection{Radial temperature distribution and mass accretion rate estimate}

The simplest way of testing theoretical disc models is to convert the
intensities in the eclipse maps to blackbody brightness temperatures,
which can then be compared to the radial run of the effective temperature
predicted by steady state, optically thick disc models. However, as
discussed by Baptista et~al. (1998), a relation between the effective
temperature and a monochromatic brightness temperature is non-trivial,
and can only be properly obtained by constructing self-consistent models
of the vertical structure of the disc. Therefore, our analysis here
is meant as preliminary, and should be complemented by detailed disc
spectrum modeling in a future paper.

\begin{figure}
\centerline{\psfig{figure=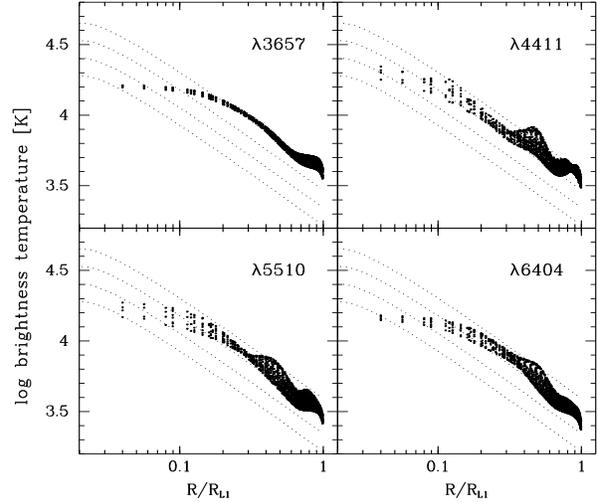,angle=-90,width=11cm,rheight=7.5cm}}
\caption{ Brightness temperature radial distributions of the UU Aqr accretion
	disc for the continuum maps of figure\,\ref{fig3}, calculated assuming
	a distance of 200 pc to the system (BSH96). Dotted lines correspond to
	steady-state disc models for mass accretion rates of \.{M}$= 10^{-8.5}$,
	$10^{-9}$, $10^{-9.5}$ and $10^{-10}\; M_\odot \; yr^{-1}$, assuming
	M$_1= 0.67 \; M_\odot$ and $R_1= 0.012 \; R_\odot$ (BSC94). Abscissae
	are in units of the distance from disc centre to the inner Lagrangian
	point (R$_{\rm L1}$). }
\label{fig6}
\end{figure}

Fig.\,\ref{fig6} shows brightness temperature radial distributions for
the continuum maps of Fig.\,\ref{fig3} in a logarithmic scale.
Each temperature shown is the blackbody brightness temperature that
reproduces the observed surface brightness at the corresponding pixel
assuming a distance of 200~pc to UU~Aqr (BSH96). Steady-state disc models
for mass accretion rates of $10^{-8.5}$, $10^{-9}$, $10^{-9.5}$ and
$10^{-10}\; M_\odot \; yr^{-1}$ are plotted as dotted lines for comparison.
These models assume M$_1= 0.67 \; M_\odot$ and $R_1= 0.012 \; R_\odot$
(BSC94).

The distributions resemble those obtained by BSH96 for the high brightness
state of UU Aqr, closely following the $T\propto R^{-3/4}$ law for steady
accretion in the intermediate and outer disc regions ($R \geq 0.2\; 
R_{\rm L1}$) but displaying a noticeable flattening in the inner disc
($R < 0.1\; R_{\rm L1}$).
Temperatures range from $\sim 18000$ K in the inner disc to 6000 K in the
outer disc regions, leading to inferred mass accretion rates of \.{M}=
$10^{-9.0 \pm 0.3}\; M_\odot \, yr^{-1}$ at $R= 0.1\; R_{\rm L1}$ and
$10^{-8.7 \pm 0.2} \; M_\odot \, yr^{-1}$ at $R= 0.3\; R_{\rm L1}$ --- in
good agreement with the results of BSH96 for the high brightness state.
The quoted errors on \.{M} account for the statistical uncertainties in the
eclipse maps, obtained from the Monte Carlo procedure described in
section\,\ref{mem}, and the scatter in the temperatures of maps at
different wavelengths.
The eclipse map at $\lambda 3657$ leads to temperatures which are
systematically higher than those of the other continuum maps of Fig.\,3,
in an example of the limitations of using brightness temperatures to
estimate the mass accretion rate. This difference reflects the fact that
the Balmer jump appears in emission for the intermediate and outer disc
regions, as will be seen in section\,\ref{spectra}.
%
%

\subsection{Radial line intensity distributions}

\begin{figure}
\centerline{\psfig{figure=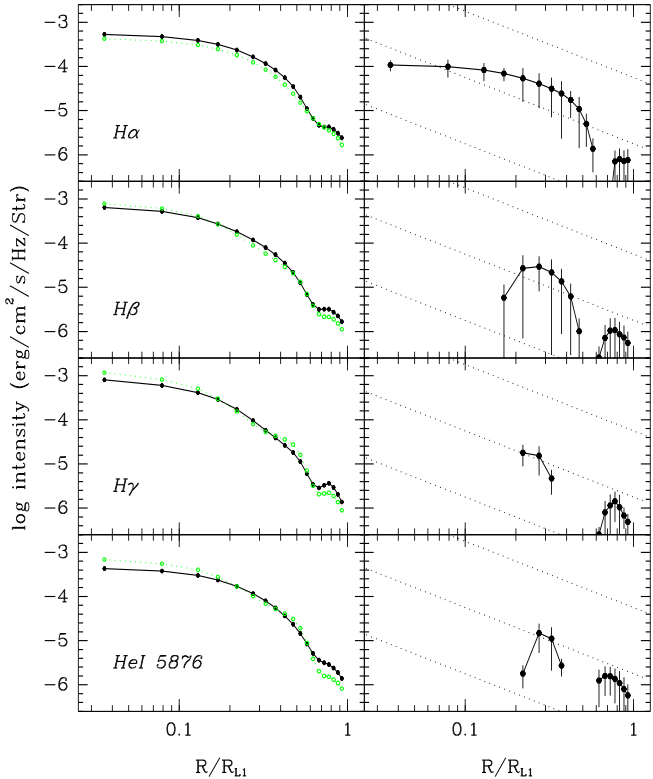,width=10cm,rheight=12cm}}
\caption{ Radial line intensity profiles for the most prominent lines,
	calculated assuming a distance of 200 pc to the system (BSH96). 
	Abscissae are in units of the distance from disc centre to the inner
	Lagrangian point (R$_{\rm L1}$). Right panels show net line emission 
	radial distributions. Dotted lines depict the slope of the expected
	relation $I \propto R^{-1.5}$. }
\label{fig7}
\end{figure}

Left panel in Fig.\,\ref{fig7} shows radial intensity distributions for
the most prominent lines (solid) and adjacent continuum (dotted) in a
logarithmic scale. The line distributions were obtained from the average
of all eclipse maps across the line region, while the continuum
distributions were obtained from the average of eclipse maps on
both sides of each line. 
Net line emission distributions were computed by subtracting the
distributions of the adjacent continuum from those of the lines,
and are shown in the right panel.
In the external map regions ($R \simgt 0.7\; R_{\rm L1}$) the intensities
of both line and continuum drop by a factor $\sim 10^3$ with respect
to the inner disc regions, making the computation of the net emission
quite noisy and unreliable.
H$\alpha$ is seen in emission (intensities larger than those at the
adjacent continuum) at all disc radii and up to $R \simeq 0.6\; R_{\rm L1}$. 
The other lines are in absorption in the inner disc and transition to
emission at intermediate ($R \sim 0.2\; R_{\rm L1}$) disc radius.
This behaviour is noticeably different from that observed at the low
brightness state, where H$\alpha$ is seen in emission in the inner disc
and disappears into the continuum for $R \simeq 0.3\; R_{\rm L1}$ (BSH96).
This result suggests that the line emission region increases in size
from the low to the high brightness state, possibly in response to changes
in mass accretion rate.
The transition from absorption to emission occurs at larger disc radii for
lines of higher excitation. This can be explained, for the Balmer lines,
by the increase in continuum emission at the inner disc for shorter
wavelengths.
%
%

A set of dotted lines in the right panel indicate the slope of the empirical
radial dependency of the line emissivity in accretion discs, $I \propto
R^{-1.5}$, as inferred from Doppler Tomography by assuming a Keplerian
distribution of velocities for the emitting gas (Marsh et al. 1990).
For H$\gamma$ and He\,I $\lambda$5876, the net emission occurs for a narrow range of radii making a comparison with the empirical law difficult.
The derived radial distributions for H$\alpha$ and H$\beta$ are clearly
different from the empirical $I \propto R^{-1.5}$ law; in particular,
the H$\alpha$ distribution is flat at inner and intermediate disc radii
($R <0.3\; R_{\rm L1}$).
This remark suggests that the line emitting regions on the disc surface
are not in Keplerian orbits or that a substantial fraction of the emission
lines does not arise from the accretion disc, in line with the inferences
drawn by the comparison of velocity-resolved eclipse maps in
section\,\ref{estrutura}.
The latter hypothesis is consistent with the significant uneclipsed
components inferred for the Balmer and He\,I lines (section
\,\ref{uneclipsed}).

\subsection{Spatially resolved spectra}
\label{spectra}

Each of the eclipse maps yields spatially-resolved information about
the emitting region on a specific wavelength range.  By combining all
narrow-band eclipse maps we are able to isolate the spectrum of the
eclipsed region at any desired position (e.g., Rutten et~al. 1994;
Baptista et~al. 1998).

To investigate the possible influence of the gas stream on the disc
emission and motivated by the observed asymmetries in the eclipse maps
shown in section\,\ref{estrutura}, we divided the disc into two major
azimuthal regions to extract spatially-resolved spectra: the gas stream
region (upper right quadrant in the eclipse maps of Figs.\,3 and 4) and
the disc region (the remaining 3/4 of the eclipse map). 
For each of these regions, we divided the maps into a set of 6 concentric
annuli centred on the white dwarf of width $0.1\: R_{L1}$ and with radius increasing in steps of $0.1\: R_{L1}$.
Each spectrum is obtained by averaging the intensity of all pixels
inside the corresponding annulus and the statistical uncertainties
affecting the average intensities are estimated with the Monte Carlo
procedure described in section\,\ref{mem}.

\subsubsection{Disc spectra}
\label{disc}

\begin{figure*}
\centerline{\psfig{figure=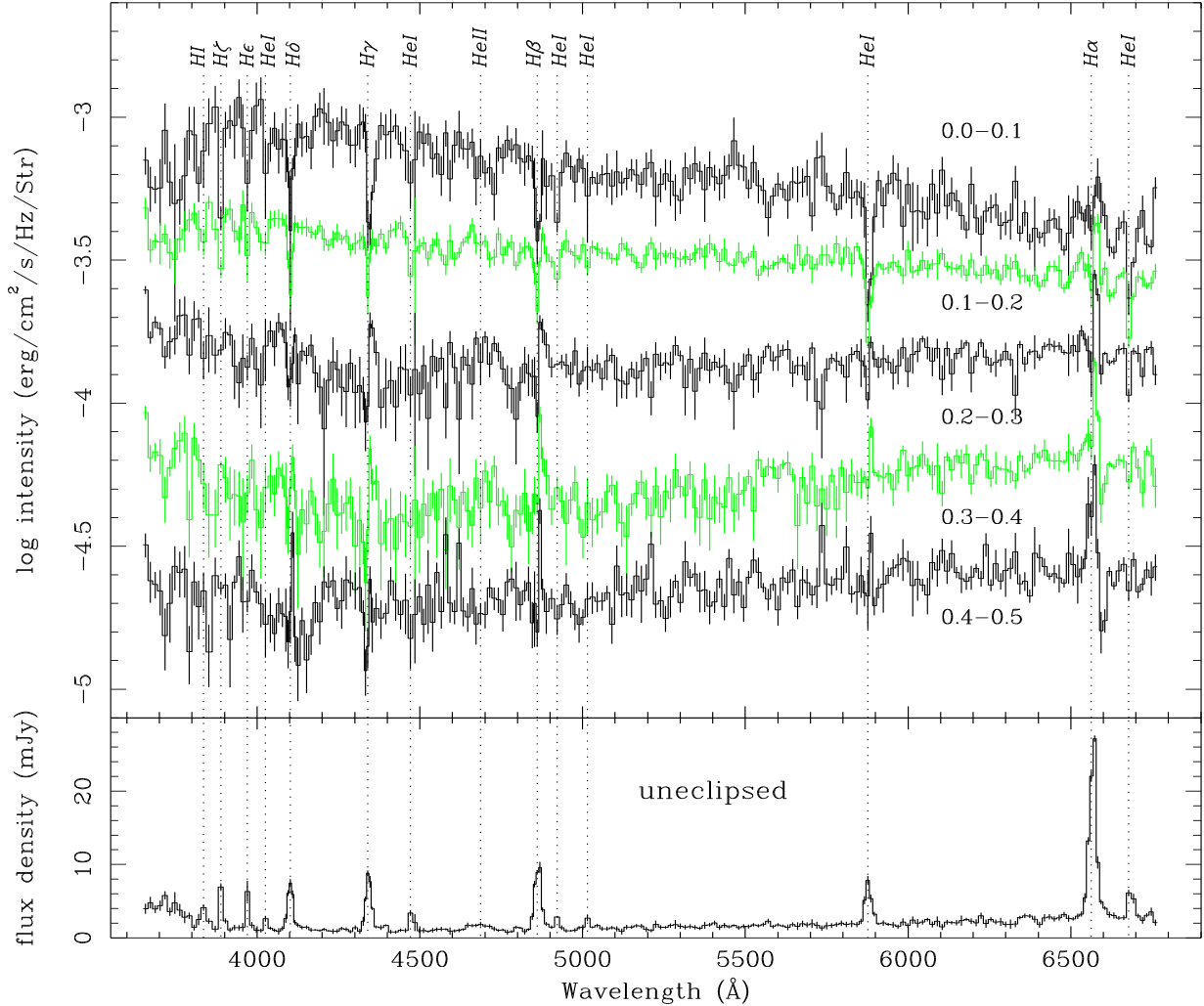,angle=-90,width=21cm,rheight=15cm}}
	\caption{ Spatially resolved spectra of the UU Aqr accretion disc.
	The spectra were computed for a set of concentric annular sections
	(radius range indicated on the right, in units of $R_{L1}$).
	The lower panel shows the spectrum of the uneclipsed light.
	The most prominent line transitions are indicated by vertical dotted
	lines.  Error bars were derived via Monte Carlo simulations with the
	eclipse lightcurves. }
\label{fig8}
\end{figure*}

Fig.\,\ref{fig8} shows spatially-resolved spectra of the disc region in
a logarithmic scale. The inner annular region is at the top and each
spectrum is at its true intensity level. The spectrum of the uneclipsed
component is shown in the lower panel and will be discussed in detail in
section\,\ref{uneclipsed}.
%
%
The spectrum of the inner disc is characterized by a blue and bright
continuum filled with deep and narrow absorption lines. The continuum
emission becomes progressively fainter and redder for increasing disc radius
while the lines transition from absorption to emission showing clear P~Cygni
profiles on all lines mapped at higher spectral resolution. The Balmer jump
appears in absorption in the inner disc and weakly in emission in the
intermediate and outer disc regions suggesting that the outer disc in
UU~Aqr is optically thin. The change in the slope and intensity
of the continuum with increasing disc radius reflects the temperature
gradient in the accretion disc, with the effective temperature decreasing
outwards.

\begin{figure}[h]
\centerline{\psfig{figure=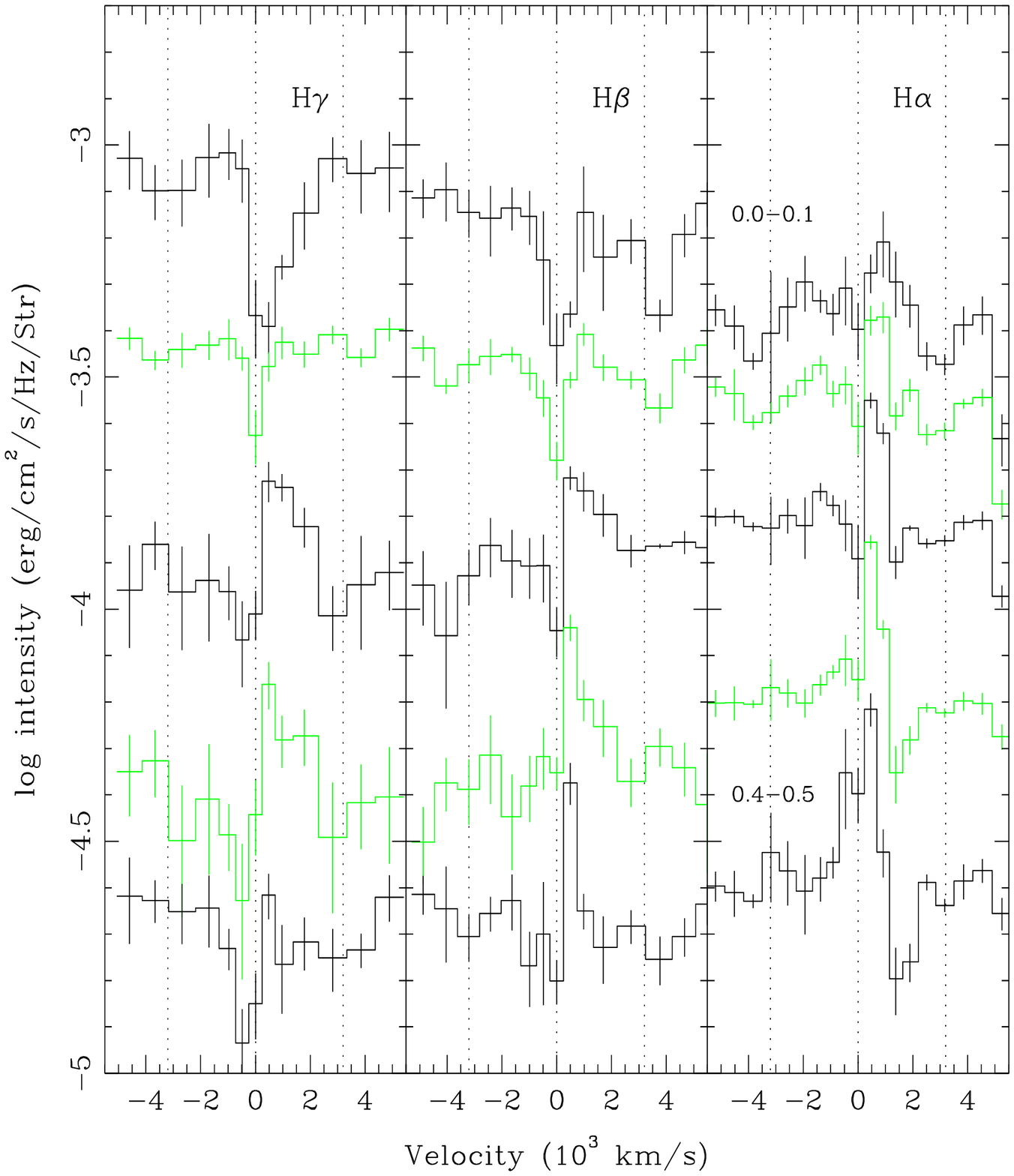,width=9.5cm,rheight=11cm}}
	\caption{ Spatially resolved spectra in the H$\alpha$, H$\beta$ and
	H$\gamma$ regions as a function of velocity. The notation is the same
	as in figure\,\ref{fig8}. Dotted vertical lines mark line centre and
	the maximum blueshift/redshift velocity expected for gas in Keplerian
	orbits around a $0.67 M_\odot$ white dwarf seen at an inclination of
	$i=78\degr$ ($v \sin i= 3200 \;km \;s^{-1}$). }
\label{fig9}
\end{figure}

The spatially resolved spectra of the disc are plotted in
Fig.\,\ref{fig9} as a function of velocity for the H$\alpha$, H$\beta$
and H$\gamma$ regions. Vertical dotted lines mark line centre and the
maximum blueshift/redshift velocity expected for gas in Keplerian
orbits around a $0.67 M_\odot$ white dwarf as seen from an inclination of
$i=78\degr$ ($v \sin i= 3200 \;km \;s^{-1}$) [BSC94].
%
%
The absorption lines at disc centre are perceptibly narrower than expected
for emission from either the white dwarf atmosphere or from disc gas in
Keplerian orbits around the white dwarf. The discrepancy increases if the
larger mass estimates of Diaz \& Steiner (1991) and Kaitchuck et~al (1998)
are assumed for the white dwarf. Moreover, the absorption lines at disc
centre are deep, while lines produced in a white dwarf atmosphere or
innermost disc regions should be broad and shallow. The width of the lines
indicate a velocity dispersion of $\simeq 1500 \;km \;s^{-1}$ for the line
emitting region in the line of sight to the disc centre and higher
velocities ($\sim 2000 \;km \;s^{-1}$) for the gas in the outer disc at
$R \simeq 0.5\; R_{\rm L1}$.
This is in clear disagreement with the expected behaviour of line emission
from gas in a Keplerian disc and provide additional evidence that these
lines do not arise from the disc atmosphere.
On the other hand, the lines at intermediate and outer disc regions
($R \simgt 0.2\; R_{\rm L1}$) show clear P~Cygni profiles indicating origin
in an outflowing gas, probably the disc wind.

We note that the H$\alpha$ line shows a redshifted ($v \sim 1800 \;km\;
s^{-1}$) absorption component in spectra of the outer disc regions ($R >
0.3\; R_{L1}$). Comparison of disc spectra at different azimuths shows
that this absorption is produced in the front side of the disc, but an
origin in the gas stream can possibly be ruled out since the absorption
component is seen with similar strengths in the leading and trailing
(the one containing the gas stream) quadrants.
The interpretation of this feature is not straightforward and deserves a
bit of caution, since it is not clearly seen in any other line and also
because the surface brightness in the corresponding disc region is only
a few percent of the intensities in the inner disc.

\subsubsection{The uneclipsed spectrum}
\label{uneclipsed}

The spectrum of the uneclipsed light (lower panel of Fig.\,\ref{fig8})
show prominent Balmer and He\,I emission lines. The Balmer jump is clearly
in emission and the optical continuum rises towards longer wavelengths
suggesting that the Paschen jump is also in emission. These results
are consistent with the findings of BSH96 and indicate that the uneclipsed
light has an important contribution from optically thin gas from outside
the orbital plane. The Balmer lines mapped at higher spectral resolution
show broad asymmetric profiles, with line peaks displaced to the red
side and wings extending up to $\simeq 1500 \;km \;s^{-1}$. The observed
asymmetry is consistent with that previously seen in the integrated spectra
of Diaz \& Steiner (1991) and Hessman (1990) and is similar to that
observed in the resonant ultraviolet lines of UX~UMa, where the uneclipsed
component was attributed to emission in a vertically-extended disc wind
(Baptista et~al. 1995; 1998; Knigge \& Drew 1997).

\begin{figure}[h]
\centerline{\psfig{figure=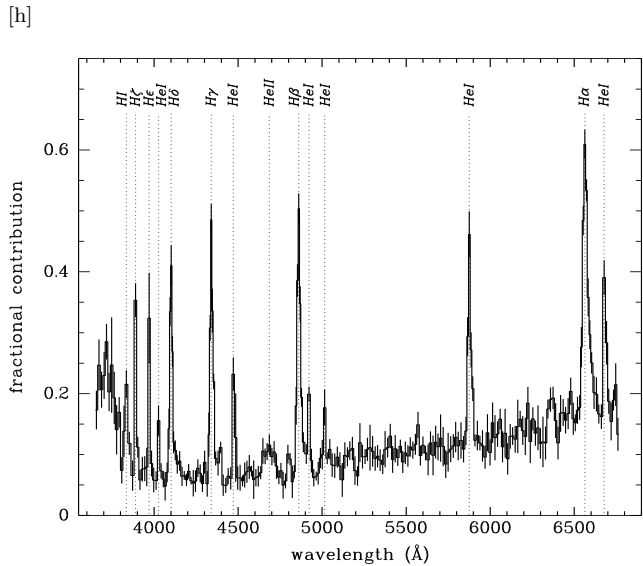,angle=-90,width=10cm,rheight=7.5cm}}
	\caption{ Fractional contribution of the uneclipsed component to
	the total flux as a function of wavelength. The values were obtained
	by dividing the flux of the uneclipsed component by the average out
	of eclipse level for each passband. }
\label{fig10}
\end{figure}

The fractional contribution of the uneclipsed component to the total flux
was obtained by dividing the flux of the uneclipsed light by the average
out of eclipse level at each passband. The result is shown in 
Fig.\,\ref{fig10}. The fractional contribution of the uneclipsed light
is very significant for the optical emission lines, reaching 40-60\ per cent
at the Balmer lines and 20-40\ per cent at the He\,I lines, and decreases
steadily along the Balmer series. 
The difference in fractional contribution between the Balmer
and He\,I lines and among the Balmer lines indicates the existence of
a vertical temperature gradient in the material above/below the disc,
with the He\,I lines (which require higher excitation energies)
being produced closer to the orbital plane. In any case, a substantial 
fraction of the light at these lines does not arise from the orbital
plane and is not occulted during eclipse.
The uneclipsed component gives significant contribution also to the
continuum emission. About 20\ per cent of the flux at the Balmer continuum
and similar fraction of the continuum emission at the red end
of the spectrum arise from regions outside the orbital plane.
%
%

\subsubsection{The gas stream region}
\label{stream}

\begin{figure*}
\centerline{\psfig{figure=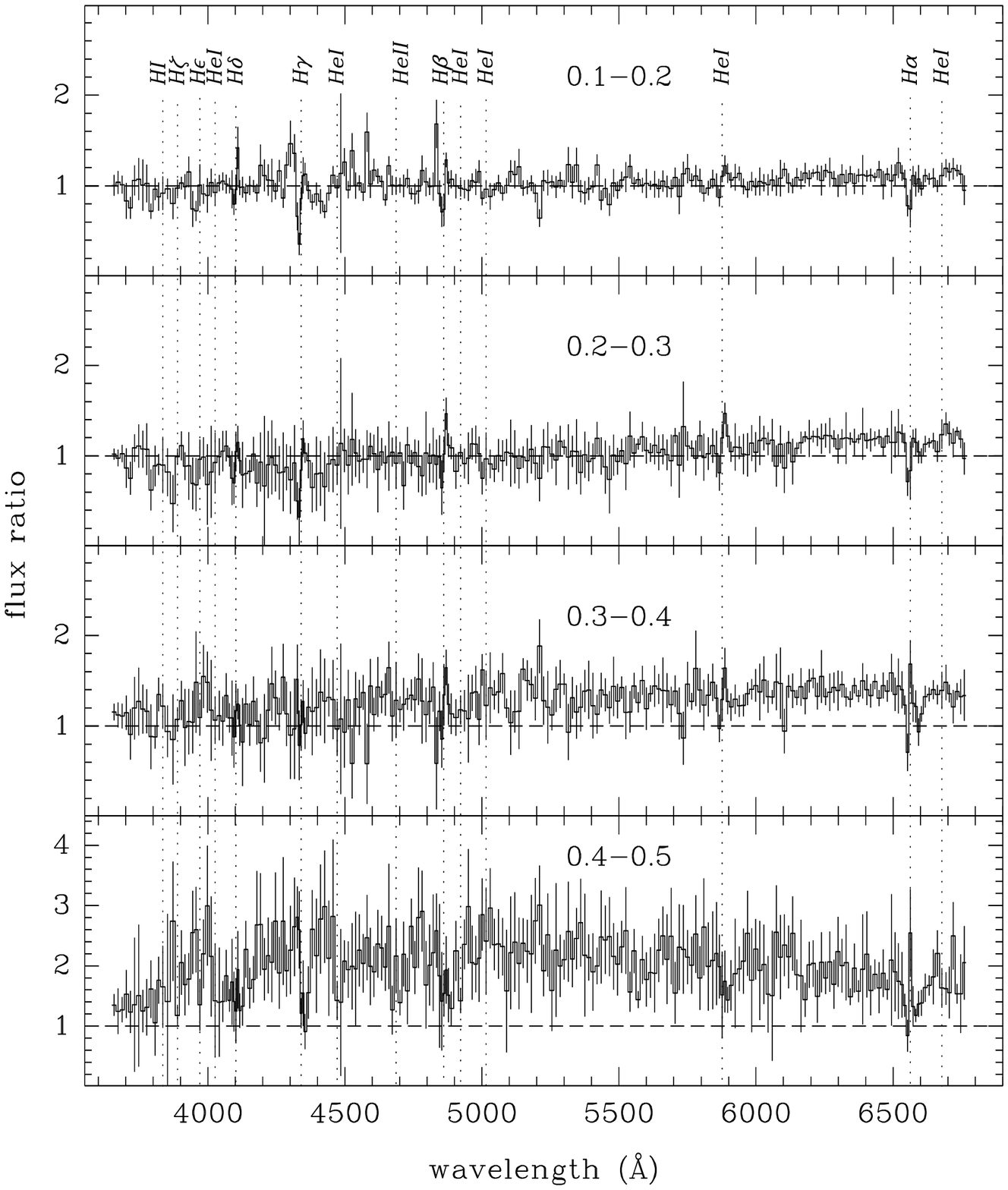,width=14cm,rheight=17cm}}
	\caption{ Ratio of the gas stream spectrum and the disc spectrum 
	for the same set of annular regions of figure\,\ref{fig8}. Dashed
	lines mark the unity level for each panel. The notation is the same
	as in fig.\,\ref{fig8}.}
\label{fig11}
\end{figure*}

Fig.\,\ref{fig11} shows the ratio between the spectrum of the gas stream
region and the disc region at same radius as a function of radius.
A dotted line marks the unity level for each panel. 
The comparison shows that the spectrum of the gas stream is noticeably
different from the disc spectrum in the outer disc regions (where one
expects a bright spot to form due to the shock between the inflowing
stream and the outer disc rim), but also reveals systematic differences
between stream and disc spectra in a range of radii. In all cases,
the stream emission is stronger than that of the adjacent disc.
%
%

This result suggests that the material in the gas stream continues to flow
downstream beyond the bright spot position. In this regard, there are three
potential cases: (1) stream overflow that arcs high above/below the disc
until it hits the disc surface again at a downstream position closer to 
the white dwarf (hereafter called the ``classical stream overflow''); 
(2) stream overflow that continuously skims the disc surface (hereafter
called the ``disc-skimming overflow''); and (3) the stream drills into the
disc at the impact site (hereafter named the ``disc stream penetration'').
The latter case seems physically unrealistic and is not supported by hydrodynamic calculations of stream-disc interaction (Armitage \& Livio
1996, 1998). 
The fact that there is enhanced emission in the stream region extending
all the way from the outer disc down to $R\simeq 0.2\; R_{L1}$ argues in
favor of an interpretation in terms of disc-skimming overflow instead of
the classical stream overflow -- as has been suggested to explain the
behaviour of SW~Sex stars (Hellier \& Robinson 1994; Hellier 1996) --
since the latter would produce enhanced emission only at the position of
the two spots, at the initial impact site in the outer disc edge and at
the re-impact site much closer to disc centre (Lubow 1989).

The spectrum of the ratio becomes redder for decreasing disc radius,
possibly a combination of the disc emission becoming bluer as one moves
inwards and the gas stream emission becoming redder while its energy is
continuously lost in the shock with disc material along the inward stream
trajectory.
This is reminiscent of what was seen in ultraviolet eclipse observations
of the dwarf novae IP~Peg in quiescence, which revealed a compact blue
bright spot with an extended red tail (Baptista et~al. 1993).

\section{Discussion} \label{discussao}

In this section we present and discuss some possible interpretations for
the results of section\,\ref{resultados} in the context of the current
models for the SW Sex stars.

\subsection{Where do the lines come from?}

In previous sections we have accumulated evidences that the behaviour
of the UU Aqr lines in its high state is not consistent with emission
in a disc atmosphere, namely: 
(i) negligible rotational disturbance, (ii) no mirror symmetry between
eclipse maps in symmetric velocity bins; (iii) H$\alpha$ line emission
distribution much flatter than the empirical $I \propto R^{-1.5}$ law;
(iv) significant uneclipsed components, and (v) presence of P~Cygni
profiles in the disc spectra at intermediate and large disc radii.
If the lines do not arise in the disc atmosphere, where do they come from?

The most compelling interpretation is that the lines are produced in a
disc chromosphere + wind. This region is hot, dense, opaque and has low
expansion velocities close to the orbital plane in order to produce the
observed deep, narrow absorption lines in the line of sight to the inner
disc. Most of the high excitation lines are produced close to the disc
plane. The density and temperature decrease with height above/below the
disc as the outflowing gas spreads over an increasing surface area.
Optically thin emission from this extended region is
probably responsible for the Balmer jump (and lines) in emission observed
in the uneclipsed spectrum.  Support in favor of this scenario comes from
the recent detailed modeling of the C\,{\sc IV} wind line of eclipsing
nova-likes by Schlosman, Vitello \& Mauche (1996) and Knigge \& Drew
(1997). Their results suggest the existence of a relatively dense ($n_e
\sim 4 \times 10^{12}$~cm$^{-3}$) and vertically extended chromosphere
between the disc surface and the fast-moving parts of the wind, which
could produce significant amounts of optically thin emission.
At orbital phases around eclipse, gas outflowing in the direction of the
secondary star will be seen along the line of sight to the bright underlying
accretion disc as blueshifted absorption features, while gas expelled in
the direction away from the secondary star should contribute with
redshifted emission.

We tested this scenario by 
comparing spatially resolved spectra of the disc lune closest to the
secondary star (the right hemisphere of the disc in the eclipse maps of
Fig.\,\ref{fig3}, hereafter called the ``front'' side) and of the disc
lune farthest away from the secondary star (the left hemisphere of the
disc in Fig.\,\ref{fig3}, hereafter called the ``back'' side). 
For this purpose, we defined two opposite azimuthal disc regions of width
$30\degr$ along the major axis of the binary, and extracted spatially
resolved spectra for the same set of annuli as above. These spatially
resolved spectra are noisier than those of Figs.\,\ref{fig8} and
\ref{fig9} because in this case the average intensity of each annulus
is computed from a significantly smaller number of pixels. The results
are shown in Fig.\,\ref{fig9A} for the H$\beta$ and H$\gamma$ regions
and are consistent with our interpretation:
The blueshifted absorption component is seen mainly in the front side of
the disc while the redshifted emission is generally more prominent in
spectra of the back side of the disc.
The fact that the blueshifted absorption can still be seen projected along
the line of sight at the outer regions of the disc favours a more spherical
or equatorial geometry for the outflowing gas instead of a highly collimated,
polar jet.
%
%
\begin{figure}
\vspace*{7cm}
	\caption{ Comparison of spatially resolved spectra of the front and
	back side of the disc (see text) in the H$\beta$ and H$\gamma$ regions.
	The notation is the same as in figure\,\ref{fig9}. For clarity, only
	the H$\gamma$ spectra of two annuli are shown. }
\label{fig9A}
\end{figure}

The chromosphere + disc wind interpretation satisfactorily accounts for all
the features listed above and also gives a plausible explanation for 
(1) the distinct semi-amplitude of the radial velocity $K$ and systemic
velocity $\gamma$ as inferred from different emission lines and 
(2) the time dependent $K$ and $\gamma$ values (Hoard et~al. 1998). 
The centroid of lines of different excitation level will occur at different
locations in the primary lobe and will sample different velocities 
along the line of sight. With respect to (2), 
the comparison of the H$\alpha$ map of BSH96 (which corresponds to the low
brightness state) with that of Fig.\,\ref{fig4} (the high state) reveals
that in the latter the emission extends over a much larger region of the
primary lobe with a pronounced asymmetry in the stream region,
suggesting that the wind emission is variable in time and is intimately
connected with the mass accretion rate. 
This remark gives additional support to the suggestion
by Hoard et~al. (1998) that the observed time dependence of the $K$ and
$\gamma$ velocities might be due to variability of a wind component
in UU Aqr.

An alternative possibility is to consider the deep absorption lines
seen towards the line of sight to the disc centre as being produced by
absorption in a vertically extended disc rim. Although this scenario
accounts for the narrow absorption lines,
it is not able to explain the large velocities inferred from the line
width for intermediate and large disc radius nor the P~Cygni profiles.
Furthermore, it should result in a perceptible front-back asymmetry in
the disc surface brightness (namely, the back side of the disc should be
brighter) which is not seen in the eclipse maps.

Recently, Horne (1999) proposed that most of the features of the SW Sex
stars could be explained in terms of a disc-anchored magnetic propeller,
in which energy and angular momentum are extracted from the magnetic field
of the inner disc regions to fling part of the material in the gas stream
out of the binary towards the back side of the disc. 
Although this model is able to explain many of the observed features of
UU Aqr, it can only account for the observed P Cygni profiles if the gas
trapped by the inner disc magnetic field is expelled in all directions
and not only towards the back of the disc.
We note that, in this case, there is no significant difference between
the propeller and the disc wind models and, in fact, the former could
possibly work as the underlying physical mechanism driving the latter.

If disk-skimming overflow does occur, we might expect that 
dissipation of energy in the collision between the gas stream and the
disc material gives rise to a bulge extending along the stream trajectory
over and under the disc. This bulge will appear in front of the
chromosphere + wind line emitting region at the inner disc when seen
along the line of sight at orbital phases 0.5-0.9.
This may explain the phase-dependent absorption lines, observed from
phases 0.5-0.9 and with maximum at phase $\sim 0.8$ (Heafner 1989;
Hoard et~al. 1998). The enhanced line emission along the gas stream
(see Fig.\,\ref{fig4}) is possibly responsible for the phase offset
between photometric and spectroscopic conjunction
(Diaz \& Steiner 1991; Hoard et~al. 1998). 

In summary, the picture which emerges from our results is consistent
with the results from the Doppler tomography and the model proposed
for UU Aqr by Hoard et~al. (1998).

\subsection{Where has the bright spot gone?}

Although our observations correspond to the high brightness state of
UU Aqr, our eclipse maps do not show the conspicuous asymmetric structure
seen in the high state eclipse maps of BSH96 and which was interpreted
as being the bright spot.
The explanation for the `disappearance' of the bright spot may be
connected with the stunted outbursts found by Honeycutt et~al. (1998).

BSH96 pointed out that the inferred accretion rate of UU Aqr
is close to the critical mass accretion rate for disc instability
to occur and remarked that the long-term lightcurves of accretion
discs with mass transfer rates near their critical limit might display
low-amplitude ($\simlt 1.0$ mag) outbursts caused by thermal
instabilities in the outer disc regions (e.g., Lin, Papaloizou \&
Faulkner 1985). In this case the outburst is restricted to the outer
1/3 of the disc extent while the inner disc remains in a high viscosity,
steady state.
Honeycutt et~al (1998) suggested that such dwarf-nova type instabilities
could be an explanation for the stunted outbursts of UU Aqr if a
mechanism can be identified to make the amplitudes appear small. 
We note that the observed low amplitudes can be easily accounted for
by the reduced contrast of the light from the outbursting outer
regions -- where the efficiency in transforming gravitational potential
energy in radiation is relatively low  -- in comparison to the bright,
optically thick and steady inner disc.

If the observed stunted outbursts of UU Aqr are caused by thermal
instabilities in its outer disc, the disc radius is expected to
increase during the outburst and will eventually reach the 3:1 tidal
resonance radius leading to an elliptical precessing disc reminiscent
of what possibly happens in SU UMa stars in superoutburst
(e.g., Warner 1995 and references therein). 
We suggest that the azimuthally elongated structure seen in the
eclipse maps of BSH96 is the signature of such an elliptical disc
and not the bright spot. Following this line of reasoning, this
structure should not be present when the disc radius is smaller than
the tidal resonance radius. Support for this interpretation comes from
the comparison of disc radius in the high state eclipse maps of BSH96
and our eclipse maps. 
From BSH96 data we estimate a disc radius of $R_d \simeq 0.7 \; 
R_{\rm L1}$, comparable to the 3:1 tidal resonance radius for a
mass ratio of $q=0.3$. Our eclipse maps lead to a smaller value of
$R_d = 0.65 \; R_{\rm L1}$. Therefore, we suggest that UU Aqr was in an 
occasional superhumper state during the high brightness state observations
of BSC94.

In the model of Hoard et~al. (1998), after the explosive impact of the
high \.{M} accretion stream with the edge of the disc, the incomming gas 
forms an optically thick absorbing bulge on the disc that either follows 
roughly the stream trajectory or runs along the rim of the disc, producing
the absorption features seen at phases 0.4-0.9. It may alternatively be
possible that the structure seen in the eclipse maps of BSH96 is the
signature of such post-impact stream material running along the edge of
the disc. In this scenario, the azimuthally extended bulge would be present
or not depending on the (variable) mass accretion rate and the resulting
orbital hump would remain fixed in phase.

It would be interesting (although outside the scope of this paper) to
reanalize the data of BSC94 to see if the orbital hump present in the high
state precesses in phase in a similar manner as superhumps in superoutbursts
(supporting the elliptical disc scenario) or if its maximum occurs always
at the same orbital phase range about 0.8 - 0.9 cycle (favouring the
post-impact bulge scenario).

\section{Conclusions} \label{conclusao}

We used time-resolved spectroscopy to study the structure and spectra
of the accretion disc and gas stream of the novalike UU~Aquarii in
the optical range. The main results of this analysis can be summarized
as follows:

\begin{itemize}

\item The spectrum of the inner disc shows a blue continuum filled with deep,
narrow absorption lines which transition to emission with clear P~Cygni
profiles at intermediate and large radii ($R\simgt 0.2 \; R_{L1}$).

\item The spectrum of the uneclipsed light has strong H\,I and He\,I
emission lines and a Balmer jump in emission indicating a significant
contribution from optically thin regions outside the orbital plane.

\item Velocity-resolved eclipse maps and spectra indicate that most
of the line emission probably arises in a vertically-extended disc
chromosphere + wind.

\item Differences in fractional contribution among emission lines suggests
a vertical temperature gradient in the material above/below the disc.

\item The comparison of the spectrum of the gas stream region and the
disc region at the same radius as a function of radius gives evidence 
of gas stream disc-skimming overflow down to $R\simeq 0.2\; R_{L1}$.
This may explain the phase-dependent absorption in emission lines.

\item The comparison of our eclipse maps with those of BSH96 suggests that
the asymmetric structure in the outer disc previously identified as
the bright spot may be the signature of an elliptical precessing disc
similar to those possibly present in SU UMa stars during superoutbursts.

\end{itemize}

\section*{Acknowledgments}

We gratefully acknowledge the director of KPNO for granting telescope time
for this project at the Summer Queue Program, Tod Boroson and the team
of observers at KPNO for their kind effort in collecting the data, 
Knox Long and the director of STScI for financial support through
the Director Discretionary fund, Susan Keener for helping with the data
reduction at STScI, and an anonymous referee for valuable comments and
suggestions that helped to improve the presentation of the results.
RB acknowledges financial support from CNPq/Brazil through grant no. 300\,354/96-7. This work was partially supported by PRONEX grant
FAURGS/FINEP 7697.1003.00.

\bsp
\end{document}